\documentclass[manuscript]{aastex}

\slugcomment{Version: 2019 August 30}

\shorttitle{Comet C/2019 J2}
\shortauthors{Jewitt and Luu}

\begin{document}


\title{Disintegrating In-Bound Long-Period Comet  C/2019 J2}


\author{David Jewitt$^{1,2}$ and Jane Luu$^{3,4}$}
\affil{$^1$ Department of Earth, Planetary and Space Sciences,
UCLA, 
595 Charles Young Drive East, 
Los Angeles, CA 90095-1567\\
$^2$ Dept.~of Physics and Astronomy,
UCLA, 
430 Portola Plaza, Box 951547,
Los Angeles, CA 90095-1547\\
$^3$ Draper Laboratory, 555 Technology Square, Cambridge, MA 02139 \\
$^4$ Centre for Earth Evolution and Dynamics, University of Oslo, Postboks 1028 Blindern, 0315 Oslo, Norway \\
}


\email{jewitt@ucla.edu}

\begin{abstract}
We present observations of the disintegrating long-period comet C/2019 J2 (Palomar) taken to determine the nature of the object and the cause of its demise.  The data are consistent with break-up of a sub-kilometer nucleus into a debris cloud of mass $\sim10^9$ kg, peaking on UT 2019 May 24$\pm$12. This is $\sim$56 days before perihelion and at a heliocentric distance of $\sim$1.9 AU.  We consider potential mechanisms of disintegration.  Tidal disruption is ruled-out, because the comet has not passed within the Roche sphere of any planet.  Impact disruption is implausible, because the comet orbit is highly inclined (inclination 105.1\degr) and disruption occurred far above the ecliptic, where asteroids are rare.  The back-pressure generated by sublimation (0.02 to 0.4 N m$^{-2}$) is orders of magnitude smaller than the reported compressive strength (30 to 150 N m$^{-2}$) of cometary material and, therefore, is of no importance.  The depletion of volatiles by sublimation occurs too slowly to render the nucleus inactive on the timescale of infall. However, we find that the e-folding timescale for spin-up of the nucleus by the action of sublimation torques is shorter than the infall time, provided the nucleus radius is $r_n <$ 0.4 km.  Thus, the disintegration of C/2019 J2 is tentatively interpreted as the rotational disruption of a sub-kilometer nucleus caused by outgassing torques.  
\end{abstract}


\section{INTRODUCTION}
Some comets  spontaneously disintegrate, for reasons which remain poorly understood (Sekanina 1984).  Disintegration  competes with devolatilization, impact into the planets or the Sun, and ejection from the solar system as one of  the leading causes of cometary demise (Jewitt 2004), but the relative rates of these processes are largely unknown.  Unfortunately, disintegrations occur spontaneously and are short-lived, making them difficult to observe.  The recent disruption of newly discovered long-period comet C/2019 J2 (Palomar) (hereafter ``J2''; Ye et al.~2019a) provides an opportunity for study.  The comet has semimajor axis $a$ = 590.1 AU, eccentricity $e$ = 0.997 and inclination $i$ = 105.1\degr, with perihelion at $q$ = 1.726 AU occurring on UT 2019 July 19.    We present imaging observations taken to assess the nature of the object and the mechanism behind its disintegration.

\section{OBSERVATIONS}
Observations were obtained using the Nordic Optical Telescope (NOT) located in the Canary Islands.  The NOT is a 2.56 m diameter telescope which, when used with the 2048$\times$2048 pixel ALFOSC camera, gives a 6.5$\times$6.5\arcmin~field of view with 0.214\arcsec~pixels.  We  obtained images on two dates, as summarized in Table (\ref{geometry}).  On July 24 we obtained three images each of 200 s duration in a Johnson/Bessel V filter (central wavelength, $\lambda_c$ = 5350\AA, full width at half maximum (FWHM) $\Delta \lambda$ = 760\AA).  On August 2 we obtained 18 images of 200 s each through the R filter ($\lambda_c$ = 6410\AA, $\Delta \lambda$ = 1480\AA).  The telescope was tracked at non-sidereal rates to follow the motion of the comet (approximate rates 3\arcsec~hour$^{-1}$ West and 153 \arcsec~hour$^{-1}$ South).  As a result, the images of field stars and galaxies are heavily trailed.  The seeing was measured from sidereally tracked images of field stars to be $\sim$1.1 arcsec FWHM on both nights.  Flat fields were constructed from images of the illuminated interior of the observatory dome.  The data were photometrically calibrated using large aperture photometry of field stars appearing in the Sloan DR14 sky survey (Blanton et al.~2017).   We transformed from the Sloan filter system to V magnitudes using the relations given by Jordi et al.~(2006) and assuming that J2 has the mean color of long-period comets (V-R = 0.47$\pm$0.02, Jewitt 2015), finding V = $g$ - 0.37, and V = $r$ + 0.26.  The transformation incurs an uncertainty of at least a few percent, which is of no consequence in the present study.

\begin{figure}
\plotone{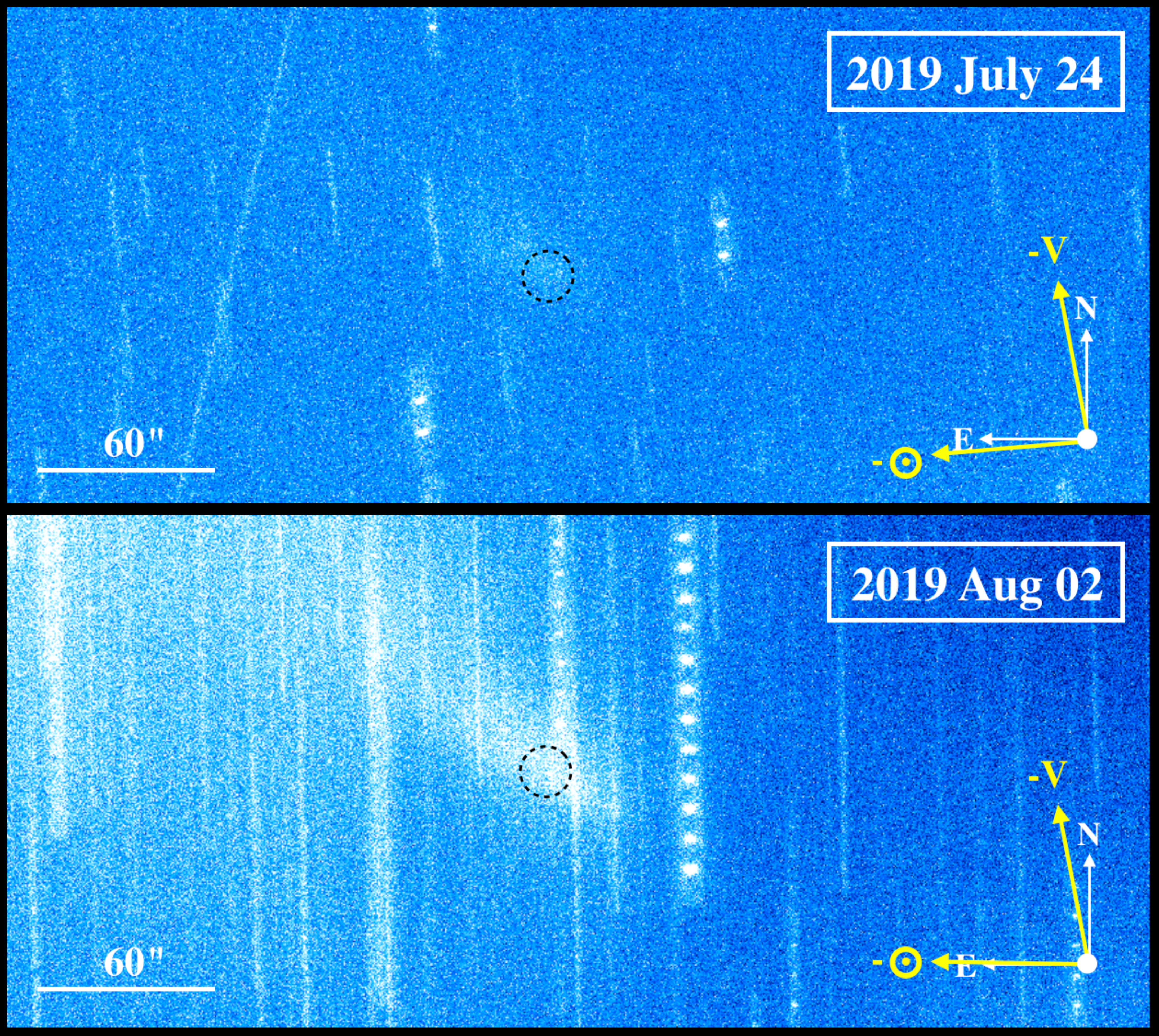}
\caption{The  structure of C/2019 J2  on (top) UT 2019 July 24 and (bottom) 2019 August 02, from the NOT telescope. The July 24 image is a  composite of three 200 s integrations through the V filter while the deeper image from Aug 02 combines 18 images of 200 s in the R band; we used a clipped median combination algorithm. The aligned bright dots are artifacts of the combination, formed by the overlapping wings of adjacent trailed field star images.  White arrows show the directions of North and East, while yellow arrows marked $-\odot$ and $-V$ show the projected anti-solar and anti-velocity vectors. The dashed, black circles show the projected 10$^4$ km radius photometry aperture and a 60\arcsec~(7.7$\times10^4$ km) scale bar is shown in each panel. \label{jul24aug02}}
\end{figure}

Composite images from the two dates are shown in Figure (\ref{jul24aug02}), with the image from August 02 being of higher signal-to-noise ratio because of the longer on-source integration time (3600 s vs.~600 s for July 24).  On both dates, the comet appears diffuse and without central condensation.  The symmetry axis of the object, while poorly defined, lies between the projected anti-solar and negative heliocentric velocity vectors, marked in the figure by yellow arrows and the $-\odot$ and $-V$ symbols, respectively. 

The diffuse appearance and low surface brightness of J2 limit the accuracy with which photometry can be obtained.  Photometry is particularly sensitive to uncertainties in the sky background, which is influenced by scattered light from stars passing through and even outside the field of view.  For example, the left-right sky gradient in the August 02 image (Figure \ref{jul24aug02}) results from light scattered from a bright object to the upper left (north east) and outside the field of view.    We extracted photometry using an aperture of projected radius 10$^4$ km on both dates (Table \ref{photometry}) to facilitate direct comparison with measurements from the Zwicky Transient Facility 1.2 m telescope, Ye et al.~(2019b).    Use of a fixed linear aperture  obviates the possibility that any changes are due to sampling different volumes in the coma.   Given the lack of a strong central concentration in the NOT data, we experimented to maximize the signal by trying different aperture positions.  The photometric uncertainties listed in Table (\ref{photometry}) are our best estimates of the errors due to the structured background and variations resulting from the uncertain position of the brightness peak.  In addition to the apparent magnitudes, the Table lists the absolute magnitudes, $H$, and the effective scattering cross-sections of the comet, $C_e$, computed as follows.   We computed $H$ from the inverse square law, expressed as

\begin{equation}
H = V - 5\log_{10}(r_h \Delta) - f(\alpha)
\label{H}
\end{equation}

\noindent where $V$ is the apparent magnitude, $r_H$ and $\Delta$ are the heliocentric and geocentric distances, respectively, and $f(\alpha)$ is the phase function.   The phase functions of comet dust are in general poorly known and, in J2, the phase function is completely unmeasured.  We used $f(\alpha) = 0.04\alpha$, which gives $B$, the ratio of scattered fluxes at 0\degr~phase and 30\degr~phase as $B$ = 3.  Values of this ratio in other comets are scattered across the approximate range 2 $< B < 3.5$, as summarized by Bertini et al.~(2019).

The absolute magnitude is related to the effective scattering cross-section, $C_e$ [km$^2$], by 
\begin{equation}
C_e = \frac{1.5\times 10^6}{p_V} 10^{-0.4 H}
\label{area}
\end{equation}

\noindent where $p_V$ is the geometric albedo.  We assume $p_V$ = 0.1, as appropriate for cometary dust (Zubko et al.~2017), but note that the albedo of J2 is unmeasured and could be higher or lower by a factor two to three.  The apparent and absolute magnitudes and the scattering cross-sections are listed  in Table (\ref{photometry}).

\section{DISCUSSION}

\subsection{Properties}

No point source nuclei are evident even in our deepest image, the 3600 s R-band integration from UT 2019 August 02 (Figure \ref{jul24aug02}, bottom panel).  The limiting magnitude of this composite is estimated as $R$ = 25.0 (3$\sigma$), which compares with $R$ = 25.5 from the ALFOSC exposure time calculator ($\url{http://www.not.iac.es/observing/forms/signal/v2.8/index.php}$) for 1.1\arcsec~seeing under dark sky conditions.  The modest difference reflects the irregular background to J2 evident in the real data.  Equation (\ref{H}) gives the corresponding limit to the absolute magnitude of any point-like nucleus as $H \ge$ 21.6 and, by Equation (\ref{area}), the maximum allowable nucleus radius is $r_n = (C_e/\pi)^{1/2} \le$ 0.1 km, again assuming $p_V$ = 0.1 (the radius limit may be scaled to other assumed albedos in proportion to $(0.1/p_V)^{1/2}$).  The imaging data thus show that no nucleus fragment larger than about 100 meters remains in early August.   

The position angle (PA) of the tail center-line in our  August 2 data is $\theta_{PA}$ = 52$\pm$5\degr.  We computed synchrones for a range of dates to find that the synchrone PA reaches the observed value for ejection on DOY 146$_{-18}^{+17}$ (UT 2019 May 26$_{-18}^{+17}$).  While the tail is too broad to be consistent with a single synchrone, the center line gives an approximate indication of the mid-time of the ejection.

\begin{figure}
\epsscale{.80}
\plotone{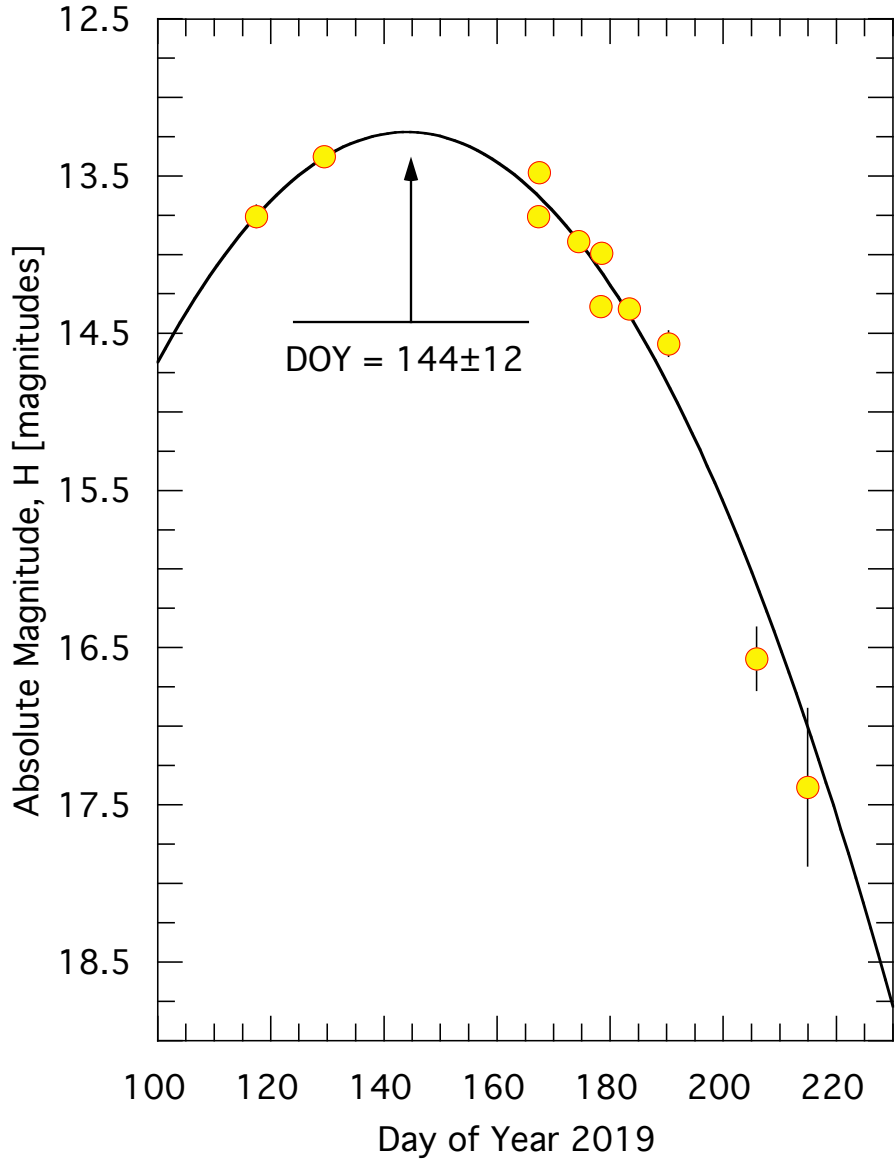}
\caption{Absolute magnitude within a circular aperture 10$^4$ km in radius, as a function of time, expressed as Day of Year (DOY = 1 on UT 2019 January 1).  The curve is a parabolic fit indicating peak $H$ at DOY = 144$\pm$12 (2019 May 24).  Data from Table (\ref{photometry}). \label{h_vs_time}}
\end{figure}

An independent estimate of the ejection timescale is obtained from the photometry listed in Table (\ref{photometry}) and plotted in Figure (\ref{h_vs_time}).  A weighted least-squares parabola fitted to the photometry gives the time of peak absolute magnitude as DOY = 144$\pm$12, corresponding to UT 2019 May 24$\pm$12.  We take this date, which is $\sim$2 months before perihelion, as our best estimate of the time of disintegration and note that the comet was then at $r_H$ = 1.87 AU.  The dates deduced independently from the tail PA and from the photometry are in agreement, within the uncertainties of measurement.  Furthermore, images posted online ($\url{http://aerith.net/comet/catalog/2019J2/pictures.html}$) show a change from centrally condensed on May 13 to diffuse with a fading core on June 4 and thereafter, consistent with a major physical change in the comet between these dates, again consistent with the inferred disruption time.

If contained in particles of mean radius $\overline{a}$, the mass of material implied by scattering cross-section $C_e$ is

\begin{equation}
M_d = \frac{4}{3} \rho \overline{a} C_e
\label{mass}
\end{equation}

\noindent where we have assumed $\rho$ = 500 kg m$^{-3}$ as the  grain density.  We obtain a crude measure of the size of the particles from the length of the dust tail, assuming that radiation pressure is the driving force.  For a constant applied acceleration, $\beta g_{\odot}$, where $\beta$ is the radiation pressure efficiency and $g_{\odot}$ is the local gravitational acceleration towards the Sun, the distance travelled by a grain released from the nucleus with zero relative speed is $L = \beta g_{\odot} \Delta t^2 / 2$, where $\Delta t$ is the time since the dust particle release.  Writing $g_{\odot} = g_1/r_H^2$, where $g_1$ = 0.006 m s$^{-2}$ is the acceleration at $r_H$ = 1 AU, we obtain 

\begin{equation}
\beta = \frac{2 L r_H^2}{g_1 \Delta t^2},
\label{beta}
\end{equation}

\noindent with $r_H$ expressed in AU.    Consider the observation on UT 2019 July 24, for which $\Delta t$ = 61 days (5.3$\times10^6$ s) from May 24 and $C_e$ = 3.5$\pm$0.7 km$^2$ within the photometry aperture of radius $L = 10^7$ m.  Substitution into Equation (\ref{beta})  gives $\beta = 4\times10^{-4}$.  For dielectric spheres, the particle radius expressed in microns is  approximately equal to the reciprocal radiation pressure factor,  $a \sim \beta^{-1}$ (Bohren and Huffman 1983), giving $a \sim 2.5$ mm.   Smaller particles should have been swept out of the aperture by radiation pressure while larger ones are retained within it.  Setting $\overline{a} \ge$  2.5 mm in Equation (\ref{mass}) gives $M_d \ge 6\times10^6$ kg.  This is a minimum mass because all the particles in the aperture must be larger than 2.5 mm in order not to have been swept out by radiation pressure.  To obtain a better estimate of $M_d$ we must consider the size distribution of the ejected particles.  

Observations from other split and disintegrating comets show that the debris size distribution approximates a power-law, such that $n(a)da = \Gamma a^{-q} da$ is the number of particles with radius in the range $a$ to $a + da$, where $\Gamma$ and $q$ are constants.   The index is typically $q \sim$ 3.5 (Jewitt et al.~2016, Ishiguro et al.~2016a, Kim et al.~2017, Moreno et al.~2012) although smaller  (e.g.~$q$ = 1.7, Kleyna et al.~2019) and larger indices (e.g.~$q$ = 3.8, Ishiguro et al.~2016b) have been reported.  If the particle radius range extends from minimum $a_{min}$ to maximum $a_{max}$, the  average radius in a $q$ = 3.5 distribution is $\overline{a} = (a_{min} a_{max})^{1/2}$.  Setting $a_{min}$ = 2.5 mm (from Equation \ref{beta}) and $a_{max} =$ 100 m (from the absence of a point-source nucleus), we find $\overline{a} =$  0.5 m.  Then, Equation (\ref{mass}) gives $M_d =$  1.2$\times10^9$ kg for the debris mass in the aperture on July 24.  This is equivalent to a sphere of the same density having radius $r_n = [3 M_d/(4\pi\rho)]^{1/3}$ or $r_n \sim  10^2$ m.  The mass in power-law distributions with $q < 4$ is dominated by the largest particles in the distribution.  For example, in a $q$ = 3.5 distribution initially extending from $a_{min} = 10^{-7}$ to $a_{max} = 10^2$ m, particles larger than 2.5 mm contain 99.5\% of the total mass.  Since these larger particles have not left the 10$^4$ km photometry aperture, we can be confident that $r_n \sim 10^2$ m is a good estimate of the equivalent radius of the disrupted body, unless the fragment size distribution is much steeper ($q$ larger) than that assumed.

\subsection{Mechanisms}
Suggested mechanisms for  cometary disintegration are many and varied.  In the case of J2 at $r_H \sim$ 1.9 AU, however, some of these mechanisms can be rejected.  Tidal disruption can be rejected, for instance, because J2 was not close to any major solar system body.  Impact disruption is implausible, because the  comet was $>$1 AU above the ecliptic at the time of disintegration as a result of the highly inclined orbit of J2 ($i$ = 105.1\degr). The number density of asteroids and other bodies drops precipitously with height  above the mid-plane and the likelihood of a disruptive collision at $>$1 AU  is vanishingly small.  

Ice sublimation can potentially lead to disintegration through distinctly different processes.  We first calculated the rate of sublimation of exposed water ice using the energy balance equation and assuming  equilibrium.  At $r_H$ = 1.9 AU, the maximum rate, found at the hotspot sub-solar point on a nucleus having Bond albedo 0.05 and emissivity of unity, is $f_s = 1.0\times10^{-4}$ kg m$^{-2}$ s$^{-1}$, and the ice temperature (depressed relative to the blackbody temperature by sublimation) is $T$ = 195 K.  The gas outflow speed is approximately given by $V_{th} =$ 420 m s$^{-1}$, the thermal speed of water molecules  (molecular weight 18) at this temperature.  The resulting back-pressure on the nucleus caused by sublimation is then $\Psi = f_s V_{th} \sim$ 0.04 N m$^{-2}$.  The same calculation repeated for supervolatile CO gives $f_s = 1.2\times10^{-3}$ kg m$^{-2}$ s$^{-1}$, and a free sublimation temperature of only $T$ = 28 K, leading to $V_{th}$ = 160 m s$^{-1}$ and a back-pressure that is only slightly larger, $\Psi =$ 0.2 N m$^{-2}$.  These back-pressures can be compared to the best available estimates of the compressive strength of cometary material, set at $S$ = 30 - 150 N m$^{-2}$ in the nucleus of 67P/Churyumov-Gerasimenko by  Groussin et al.~(2019).  With $\Psi \sim 10^{-2}S$ to 10$^{-3}S$, we reject free sublimation as a likely  cause of  nucleus cracking or disintegration.

Sublimation at specific rate $f_s$ leads to recession of the sublimating surface at rate $|dr_n/dt| = (f_s/\rho)$ m s$^{-1}$.  With $f_s = 1.0\times10^{-4}$ kg m$^{-2}$ s$^{-1}$ and $\rho$ = 500 kg m$^{-3}$ we find $|dr_n/dt| = 2\times10^{-7}$ m s$^{-1}$.  Even a  very modest nucleus of radius $r_n$ = 100 m could sustain sublimation at this rate for $r_n/(|dr_n/dt|) \sim$ 10 years, showing that devolatilization on a timescale of weeks is unlikely.  Devolatilization on a timescale of months would only be possible if the ice on the nucleus of J2 were confined to a thin ($\ll 1$ m) surface skin, but this geometry seems contrived.  

A more promising mechanism is spin-up  of the nucleus by sublimation torques, potentially driving the nucleus to rotational instability.  The e-folding timescale for spin-up due to sublimation, $\tau_s$, is a strong function of nucleus radius, $r_n$, given by
  
\begin{equation}
\tau_s = \left(\frac{16\pi^2}{15}\right)  \left(\frac{\rho_n r_n^4}{k_T V_{th}  P}\right) \left(\frac{1}{\overline{\dot{M}}}\right),
\label{tau_s}
\end{equation}

\noindent where $\rho_n$ is the density, $k_T$ is the dimensionless moment arm, $V_{th}$ is the speed of the sublimated  material, $P$ is the starting rotation period of the nucleus and $\dot{M}$ is the mass loss rate. While the numerical multiplier in this equation is geometry dependent and therefore uncertain (c.f. Jewitt 1997, Samarasinha and Muller 2013), the key factor is the strong  dependence of $\tau_s$ on the nucleus radius, $\tau_s \propto r_n^4$, if all other factors are equal.  

The physical quantities in Equation (\ref{tau_s}) are not measured for J2, but we can use evidence gleaned from the study of other comets to at least consider the possibility that spin-up might be effective in this object.  Accurate measurements of cometary densities are few and far between.  Those that exist are consistent with $\rho_n$ = 500 kg m$^{-3}$ (Groussin et al.~2019).  As above, we take $V_{th}$ = 420 m s$^{-1}$.   The median dimensionless moment arm in comets is $k_T$ = 0.015 (Jewitt 2019).  We take $P$ = 10 hour (3.6$\times10^4$ s) as a nominal nucleus rotation period (Kokotanekova et al.~2018).  Jorda et al.~(2008) determined an empirical relationship between the apparent magnitude of a comet and its hydroxyl production rate, $Q_{OH}$, namely $\log(Q_{OH}) = 30.68 - 0.25 m_H$, where $m_H$ is the apparent magnitude reduced to unit geocentric distance by the inverse square law.   Taking the UT 2019 April 27 magnitude, $V$ = 18.19$\pm$0.07 (Table \ref{photometry}), we estimate $m_H$ = 16.02 and find $Q_{OH} = 10^{26.7}$ s$^{-1}$, corresponding to $\dot{M}$ = 15 kg s$^{-1}$.  Substituting into Equation (\ref{tau_s}) gives $\tau_s \sim 1\times10^9 r_n^4$ s, with $r_n$ expressed in km.  If $r_n$ = 0.1 km, we find $\tau_s \sim 10^5$ s.

To judge the importance of spin-up, we compare $\tau_s$ with the characteristic timescale for change of the heliocentric distance, given by $\tau = r_H / |\dot{r_H}|$.  We reason that pre-perihelion spin-up is likely when $\tau_s < \tau$.  At the time of the disintegration in mid-May 2019, J2 had $r_H \sim$ 1.9 AU and $|\dot{r_H}| \sim$ 10 km s$^{-1}$, giving $\tau \sim 3\times10^7$ s.  The requirement $\tau_s < \tau$ is satisfied for $r_n < 0.4$ km, meaning that J2 could have been torqued to break-up if its nucleus  was initially smaller than about 400 m in radius, consistent with the $\sim$0.1 km scale obtained above.   As another consistency check, we note that the specific sublimation rate, $f_s= 1.0\times10^{-4}$ kg m$^{-2}$ s$^{-1}$ and the Jorda-derived mass loss rate, $\dot{M}$ = 15 kg s$^{-1}$, imply a sublimating area $C = \dot{M}/f_s$ = 1.5$\times10^5$ m$^2$ (0.15 km$^2$), equal to the surface area of a sphere 0.11 km in radius.  

Again, while the available information is insufficient to prove  that rotational instability caused J2 to disintegrate, the above considerations show that the  data are consistent with this possibility.   Disintegrations have been described in several  comets having small perihelia.  Notable examples include C/1925 X1 (Ensor) with $q$ = 0.323 AU (Sekanina 1984); C/1999 S4 (LINEAR) with $q$ = 0.765 AU (Weaver et al.~2001),  C/2012 S1 (ISON) with $q$ = 0.013 AU (Keane et al.~2016) and C/2010 X1 (Elenin) with $q$ = 0.482 AU (Li and Jewitt 2016).  These objects all displayed a diffuse, elongated appearance similar to that of J2.  The  nuclei of comets ISON and Elenin had  radii $r_n \sim$ 0.5 km (Keane et al.~2016) to 0.6 km (Li and Jewitt 2016); the radius of comet LINEAR is uncertain, but estimated as 0.1 km or larger (Weaver et al.~2000), while the size of the nucleus of comet Ensor is not known.  

We suggest that the disruption of comet J2 is possible, even at a heliocentric distances as large as 2 AU, because of its diminutive nucleus and the strong radius dependence of the e-folding spin-up time (Equation \ref{tau_s}).  We  further surmise that rotational breakup of long-period nuclei may contribute to, or even account for, the ``fading problem'', i.e.~the long-recognized inability of purely dynamical models to account for the measured distribution of cometary orbital binding energies (Oort 1950, Wiegert and Tremaine 1999, Levison et al.~2002).  Lastly, we observe that the strong size-dependence of the rotational break-up e-folding time should lead to the preferential depletion of small nuclei, and to a flattening of the size distribution of small long-period comets.  While reliable measurements of sub-kilometer nuclei are few, flattening of the distribution has indeed been inferred in a study by Fern{\'a}ndez and Soza (2012).

\clearpage 

\section{SUMMARY}
We obtained observations of the in-bound, disintegrating long-period comet C/2019 J2 (Palomar) with the 2.56 m Nordic Optical Telescope on UT 2019 July 24 and August 2.  The measured properties are consistent with the rotational disruption of a sub-kilometer nucleus under the action of outgassing torques.   Specific results include

\begin{enumerate}

\item Peak brightness was reached on UT 2019 May 24$\pm$12, when at heliocentric distance 1.9 AU and $\sim$56 days before perihelion.  We find a debris mass  $M_d \sim 1.2\times10^9$ kg, comparable to the mass of a 100 m radius sphere if density $\rho$ = 500 kg m$^{-3}$.

\item The comet appears  elongated and diffuse, with no central condensation detected down to a 3$\sigma$ limiting apparent magnitude  R = 25.0 (absolute magnitude H = 21.6). This sets a limit to the equivalent spherical radius $r_n \le$ 0.1 km (assuming geometric albedo 0.1).

\item Tidal disruption and impact disruption are rejected as likely mechanisms because the comet disrupted far from any planet and $>$1 AU above the ecliptic plane, respectively.  Neither sublimation  back-pressure nor devolatilization of the nucleus play a role in the disintegration because, at 1.9 AU from the Sun, the equilibrium sublimation rate is very small.

\end{enumerate}

\acknowledgments

We thank Yoonyoung Kim, Pedro Lacerda and the anonymous referee for helpful comments on the manuscript,  Chien-Hsiu Lee for sharing data from the NOT and John Telting and Anlaug Amanda Djupvik for help with the observations.

{\it Facilities:}  \facility{NOT}.

\clearpage

\clearpage

\clearpage

\begin{deluxetable}{lcccrccccr}
\tablecaption{Observing Geometry 
\label{geometry}}
\tablewidth{0pt}
\tablehead{ \colhead{UT Date and Time} & DOY\tablenotemark{a}   & $\Delta T_p$\tablenotemark{b} & $\nu$\tablenotemark{c} & \colhead{$r_H$\tablenotemark{d}}  & \colhead{$\Delta$\tablenotemark{e}} & \colhead{$\alpha$\tablenotemark{f}}   & \colhead{$\theta_{\odot}$\tablenotemark{g}} &   \colhead{$\theta_{-v}$\tablenotemark{h}}  & \colhead{$\delta_{\oplus}$\tablenotemark{i}}   }
\startdata

2019 July 24         22:20 - 22:27       & 205 & 5 & 2.6      &  1.728 &  1.789     &  33.5 & 94.1 & 10.9 & -32.8 \\
2019 August 02 21:51 - 23:00            &  214 & 14 &  8.1     &  1.737  & 1.860  & 32.7  & 88.1  & 11.1 & -30.4 \\
\enddata


\tablenotetext{a}{Day of Year, UT 2019 January 01 = 1}
\tablenotetext{b}{Number of days from perihelion (UT 2019-Jul-19 = DOY 200). Negative numbers indicate pre-perihelion observations.}
\tablenotetext{c}{True anomaly, in degrees}
\tablenotetext{d}{Heliocentric distance, in AU}
\tablenotetext{e}{Geocentric distance, in AU}
\tablenotetext{f}{Phase angle, in degrees}
\tablenotetext{g}{Position angle of the projected anti-Solar direction, in degrees}
\tablenotetext{h}{Position angle of the projected negative heliocentric velocity vector, in degrees}
\tablenotetext{i}{Angle of Earth above the orbital plane, in degrees}

\end{deluxetable}

\clearpage

\begin{deluxetable}{lcccccccccc}
\tablecaption{Photometry with Fixed  Radius  Apertures
\label{photometry}}
\tablewidth{0pt}
\tablehead{
\colhead{UT Date}    & \colhead{$F$\tablenotemark{a}}    & \colhead{DOY\tablenotemark{b}} &  \colhead{mag\tablenotemark{c}} & \colhead{V\tablenotemark{d}} 
&  \colhead{$r_H\tablenotemark{e}$}&  \colhead{$\Delta$\tablenotemark{f}}&  \colhead{$\alpha$\tablenotemark{g}} &  \colhead{$H$\tablenotemark{h}}&  \colhead{$C_e$\tablenotemark{i}}}

\startdata
Apr 27.49	& r	& 117.49	& 17.93$\pm$0.07	& 18.19	& 2.033	& 2.720	& 18.0	& 13.76	& 47.1$\pm$3.3  \\
May 9.47	& r	& 129.47	& 17.45$\pm$0.04	& 17.71	& 1.956	& 2.530 	& 21.5	& 13.38	& 66.8$\pm$2.8  \\
Jun 16.34	& r	& 167.34	& 17.44$\pm$0.05	& 17.70	& 1.781	& 1.943	& 31.2	& 13.76	& 47.2$\pm$2.4  \\
Jun 16.45	& g	& 167.45	& 17.79$\pm$0.05	& 17.42	& 1.781	& 1.943	& 31.2	& 13.48	& 61.0$\pm$2.3  \\
Jun 23.43	& r	& 174.43	& 17.53$\pm$0.05	& 17.79	& 1.761	& 1.862	& 32.4	& 13.91	& 40.7$\pm$2.1  \\
Jun 27.34	& r	& 178.34	& 17.91$\pm$0.05	& 18.17	& 1.751	& 1.826	& 32.9	& 14.33	& 27.8$\pm$1.4  \\
Jun 27.47	& g	& 178.47	& 18.20$\pm$0.06	& 17.83	& 1.751	& 1.826	& 32.9	& 13.99	& 38.0$\pm$1.7  \\
Jul 2.35	& r	& 183.35	& 17.89$\pm$0.05	& 18.15	& 1.741	& 1.791	& 33.4	& 14.35	& 27.4$\pm$1.4  \\
Jul 9.32	& g	& 190.32	& 18.71$\pm$0.08	& 18.34	& 1.732	& 1.761	& 33.8	& 14.57	& 22.4$\pm$1.4  \\
Jul 24.9	& V	& 205.90	& 20.36$\pm$0.20	& 20.36	& 1.728	& 1.789	& 33.5	& 16.57	& 3.5$\pm$0.7  \\
Aug 02.9	& R	& 214.90	& 21.70$\pm$0.50	& 21.23	& 1.737	& 1.860	& 32.7	  	& 17.38	& 1.7$\pm$0.4  \\

\enddata


\tablenotetext{a}{Filter used; $g$ and $r$ data are from Ye et al.~(2019b), V and R data from this work}
\tablenotetext{b}{Day of Year}
\tablenotetext{c}{Filter magnitude}
\tablenotetext{d}{Equivalent V magnitude}
\tablenotetext{e}{Heliocentric distance, in AU}
\tablenotetext{f}{Geocentric distance, in AU}
\tablenotetext{g}{Phase angle, in degrees}

\tablenotetext{h}{Absolute magnitude computed using Equation (\ref{H})}
\tablenotetext{i}{Cross-section computed from $H$ using Equation (\ref{area}) with $p_V$ = 0.1}

\end{deluxetable}


\clearpage 

%

\end{document}